\documentclass[letterpaper, conference]{IEEEtran}
\usepackage[margin=54pt]{geometry}
\usepackage{afterpage}
\IEEEoverridecommandlockouts

\usepackage[cmex10]{amsmath}
\usepackage{cite}
\usepackage[table]{xcolor}
\usepackage{makecell}

\usepackage{outlines}
\usepackage{amssymb}
\usepackage{amsfonts}
\usepackage{pifont}

\usepackage{enumitem}
\usepackage{path}
\usepackage{verbatim}
\usepackage{algorithm}
\usepackage{algorithmic}
\usepackage{framed}
\usepackage{footnote}
\usepackage{color}
\usepackage{amsmath}
\usepackage{cases}
\usepackage{subfigure}
\usepackage{theorem}
\usepackage{xurl}

\usepackage{tabularx}
\usepackage{booktabs}
\usepackage[colorlinks=true, allcolors=black]{hyperref}
\usepackage{threeparttable}

\usepackage{graphicx}

\usepackage{pgfplots}
\usepackage{pgfplotstable}
\pgfplotsset{compat=newest}
\graphicspath{{graphics/}}

{ \theorembodyfont{\normalfont} 

}

\def\QEDclosed{\mbox{\rule[0pt]{1.3ex}{1.3ex}}} 
\def\QEDopen{{\setlength{\fboxsep}{0pt}\setlength{\fboxrule}{0.2pt}\fbox{\rule[0pt]{0pt}{1.3ex}\rule[0pt]{1.3ex}{0pt}}}}
\def\QED{\QEDopen}

\newcounter{enumctr}

\DeclareFontFamily{U}{mathx}{\hyphenchar\font45}
\DeclareFontShape{U}{mathx}{m}{n}{<-> mathx10}{}
\DeclareSymbolFont{mathx}{U}{mathx}{m}{n}
\DeclareMathAccent{\widebar}{0}{mathx}{"73}

\newcolumntype{C}{>{\centering\arraybackslash}X} 
\begin{document}
    \newgeometry{top=72pt,bottom=54pt,left=54pt,right=54pt}
    
    \title{\LARGE Breathing Green: Maximising Health and Environmental Benefits for Active Transportation Users Leveraging Large Scale Air Quality Data}

    \author{Sen Yan, Shaoshu Zhu, Jaime B. Fernandez, Eric Arazo Sánchez, \\Yingqi Gu, Noel E. O’Connor, David O’Connor and Mingming Liu 
        \thanks{S. Yan, S. Zhu, J.B. Fernandez, E. Arazo, M. Liu and N. E. O'Connor are with the School of Electronic Engineering and SFI Insight Centre for Data Analytics at Dublin City University, Ireland. Y. Gu is with Alkermes plc, Dublin, Ireland. D. O'Connor is with the School of Chemical Sciences at Dublin City University. This project is now available on GitHub and can be accessed via {\tt \url{ https://github.com/SFIEssential/BreathGreen}}. Corresponding author's email: {\tt mingming.liu@dcu.ie}}
    }
    
    \maketitle

    \begin{abstract}
    \label{sec:abstract}
    Pollution in urban areas can have significant adverse effects on the health and well-being of citizens, with traffic-related air pollution being a major concern in many cities. Pollutants emitted by vehicles, such as nitrogen oxides, carbon monoxide, and particulate matter, can cause respiratory and cardiovascular problems, particularly for vulnerable road users like pedestrians and cyclists. Furthermore, recent research has indicated that individuals living in more polluted areas are at a greater risk of developing chronic illnesses such as asthma, allergies, and cancer. Addressing these problems is crucial to protecting public health and maximising environmental benefits. In this project, we explore the feasibility of tackling this challenge by leveraging big data analysis and data-driven methods. Specifically, we investigate the recently released Google Air Quality dataset and devise an optimisation strategy to suggest green travel routes for different types of active transportation users in Dublin. To demonstrate our achievement, we have developed a prototype and have shown that citizens who use our model to plan their outdoor activities can benefit notably, with a significant decrease of 17.87\% on average in pollutant intake, from the environmental advantages it offers.
    \end{abstract}


    \section{Introduction}
    \label{sec: intro}
    The detrimental effects of air pollution, specifically nitrogen oxides (${\rm NO_x}$) and particulate matter (${\rm PM}$), on human health have emerged as a significant and escalating concern. These pollutants are predominantly emitted from diverse sources such as vehicle exhaust, industrial activities, and residential heating. The well-documented consequences on human health are substantial, resulting in a considerable number of excess deaths globally (7 million) and in Ireland (3300) annually. 

${\rm NO_x}$ refers to a group of gases that includes nitrogen oxide (${\rm NO}$) and nitrogen dioxide (${\rm NO_2}$), both of which are significant contributors to air pollution. These gases have the potential to irritate the respiratory system and worsen pre-existing respiratory conditions, such as asthma, which is a prevalent health issue in Ireland (ranking fourth highest in the world). They can also increase the susceptibility to respiratory infections. Furthermore, long-term exposure to ${\rm NO_x}$ has been associated with the development of cardiovascular diseases, including heart attacks and strokes. The elevated concentrations of ${\rm NO_x}$ in urban areas, particularly in proximity to busy roads and industrial zones, pose a greater health risk to individuals, especially vulnerable groups such as the young, elderly, and those with preexisting health conditions.

Particulate matter (${\rm PM}$) refers to small particles suspended in the air, which are classified into two main categories based on their size: ${\rm PM_{10}}$ (particles with a diameter of 10 micrometres or less) and ${\rm PM_{2.5}}$ (particles with a diameter of 2.5 micrometres or less). These particles have the ability to penetrate deeply into the respiratory system when inhaled, leading to respiratory issues such as coughing, wheezing, and shortness of breath. Moreover, they can have detrimental effects on cardiovascular health, including an increased risk of heart attacks and irregular heartbeats. Additionally, exposure to ${\rm PM}$ has been associated with the development of lung cancer and adverse birth outcomes.

In Ireland, ${\rm NO_x}$ and ${\rm PM}$ emissions originate from various sources. Road transportation, especially vehicles powered by diesel engines, constitutes a significant contributor to ${\rm NO_x}$ emissions. The combustion of solid fuels, such as coal and peat, for residential heating purposes, is another prominent source of ${\rm PM}$ emissions. Industrial activities encompassing power generation and manufacturing processes also contribute significantly to the release of both ${\rm NO_x}$ and ${\rm PM}$ pollutants into the atmosphere. It is crucial to address these diverse emission sources and implement effective measures to mitigate their impact on air quality and public health.

Mitigating exposure to air pollution can have significant benefits within Ireland, as the health concerns associated with it impose a cost of approximately €2.3 billion on the Irish State annually. Studies focusing on the environmental and public health implications in the transportation sector have underscored the urgent need to reduce greenhouse gas emissions and suspended particulate matter. Notably, research has been conducted to assess the levels of traffic-related pollutants on various school commuting routes in Canada \cite{Elford2019} and China \cite{An2022}, emphasising the importance of mitigating pollutant emissions, developing smart mobility devices that prioritise user health, and identifying more efficient routes with improved air quality. These efforts are crucial for safeguarding public health, especially for vulnerable road users.

In our research, we placed emphasis on identifying routes with minimal air pollution levels and offering corresponding recommendations to transportation users. The primary contr-

\restoregeometry

\noindent ibutions of our work can be summarised as follows:

\begin{itemize}
    \item The analysis has been conducted on the air quality data from May 2021 to August 2022 in Dublin, Ireland.
    \item Real data has been processed, which included filling missing data values, based on geographic relevance and used in experiments to examine the performance of our system in different situations.
    \item A system has been developed to recommend routes to minimise pollutant inhalation for active transportation users.
\end{itemize}

The structure of this paper is as follows. In \autoref{sec: review}, we provide a review of the relevant literature on the environmental and public health implications in the transportation field. The research question is explained in detail in \autoref{sec: state}. \autoref{sec: arch} describes the system architecture in our work, and \autoref{sec: algo_exp} describes and analyses our algorithm, the dataset and the experiments. The relevant discussions and analysis are included in \autoref{sec: res_dis}. Finally, we conclude our work in \autoref{sec: con} and discuss future plans and potential improvements.
    
    \section{Related Works} 
    \label{sec: review}
    Different approaches have been employed to address the challenge of reducing pollutant emissions, particularly in the context of route planning with an emission cap for green vehicles or automated vehicles. This topic has gained significant attention in recent research \cite{Islam2021, Abdullahi2021, Olgun2021, Cai2021}. On the other hand, studies have also explored the development of smart mobility devices aimed at minimising the impact of polluting devices on individuals. For instance, in \cite{Sweeney2019}, a cyber-physical control system was designed to regulate the interaction between cyclists and electric motors. This system effectively manages the ventilation rate of cyclists by intelligently employing electric motor assistance. Similarly, an algorithm was developed in \cite{Herrmann2018} to control the switching of hybrid vehicle motors between electric and combustion modes, aiming to protect cyclists from the harmful effects of exhaust-gas emissions. Other similar studies have also been conducted in \cite{Dong2019, Hong2021, Umezawa2022}.

Moreover, researchers have also explored methodologies for user-health-friendly path planning. In the study mentioned in \cite{Langbridge2022}, the authors propose the concept of personalised commute optimisation as a means to significantly reduce the risk of pollution exposure. They highlight the potential benefits of tailoring commute routes and modes of transportation to individual users, taking into account factors such as pollution levels, personal health conditions, and preferences. Additionally, the authors present a proof-of-concept system for optimising pollution inhalation during the commute. This system aims to provide users with personalised recommendations for travel routes and transportation options that minimize their exposure to pollutants, thereby promoting healthier and more sustainable commuting practices.

A recent study \cite{LuengoOroz2019} investigated road safety and exposure to air pollution along three routes from central Edinburgh to the science and engineering campus of the University of Edinburgh. The results revealed significant differences in ultrafine particle exposure across the three routes. Additionally, a study conducted in Celje, Slovenia, proposed an alternative route to reduce cyclist exposure to black carbon based on the analysis of air quality data \cite{Jereb2018}. In the study referenced \cite{Sweeney2017}, a cyber-physical system is proposed as a solution to mitigate the impact of urban pollution on cyclists. The system operates by indirectly controlling the breathing rate of cyclists when they traverse polluted areas. A notable study conducted in \cite{Samal2020} focused on optimising the routing plan for smart city users with a specific emphasis on reducing $PM_{10}$ levels. This was accomplished by implementing the Empirical Bayesian Kriging (EBK) model during the geospatial analysis phase. In a subsequent work \cite{Samal2021}, the authors extended their approach by incorporating convolutional neural networks to extract features and utilising long short-term memory models for capturing the sequential dependency of air quality. The results consistently demonstrated that although the predicted path might be longer than the shortest route, it effectively minimises the risk of pollution exposure. 

Furthermore, the authors devised an intelligent speed advisory system in \cite{Gu2018} that provides recommendations on a suitable speed for a group of cyclists, considering the varying fitness levels within the group or the different levels of electric assistance, particularly in cases where some or all cyclists utilize electric bikes. Similarly, the authors \cite{Wu2022} developed a random forest land-use model to map $PM_{2.5}$ concentrations by investigating the exposure among cyclists in three Asian cities. By employing spatial k-fold cross-validation, the authors found that the proposed model achieved an $R^2$ value exceeding 0.67. Notably, the study concluded that for cycling commuters, selecting alternative routes could result in a reduction of over 30\% in $PM_{2.5}$ exposure.

However, the studies mentioned earlier could benefit from the utilisation of real-world data and the consideration of multiple pollutants. In our work, we address these limitations by employing an optimisation algorithm to identify the route with the minimum air pollution levels. This is achieved by utilising real-world data that encompasses multiple pollutants, allowing users to customise the weights assigned to each pollutant based on their preferences and requirements.
    
    \section{Research Problem Statement}
    \label{sec: state}
    In this section, we describe the system model for our application. The main objective of our application is to recommend a green route to a cyclist that minimizes their exposure to a specified pollutant index while travelling from a specific starting point to a destination within a bounded geographical area. Specifically, we model the area of interest using a weighted directed graph $G = (V, E, W)$, where $V = \left\lbrace 1, \ldots, n \right\rbrace$ is the vertex set, and $E = \left\lbrace e_1, \ldots, e_m \right\rbrace$ is the set of edges with corresponding positive weights in $W = \left\lbrace w_1, \ldots, w_m \right\rbrace$. The nodes in the graph represent intersections or points of interest, while the edges indicate the road segments that connect these nodes. The weight $w_i$ associated with the edge $e_i$ represents the aggregated level of pollutant exposure that a cyclist may encounter while travelling along that edge, based on historical data measurements collected at that edge and the vulnerability of the cyclist to each pollutant being considered. To further illustrate this point, mathematically, let $K$ denote the total number of pollutants being considered, and let $p_{ij}$ be the average pollutant level for the $j$'th pollutant measured at the $i$'th edge, where $i = 1, \ldots, n$ and $j = 1, \ldots, K$, and $l_i$ be the length of the $i$'th edge, then $w_i$ can be defined as follows:
    
\begin{equation}
    w_i = l_i \sum_{j=1}^{K} \alpha_j p_{ij}
\end{equation}  

\noindent where $\alpha_1, \alpha_2, \ldots, \alpha_K$ are positive coefficients capturing the vulnerability of each pollutant to the cyclist, subject to the constraint $\sum_{j=1}^{K} \alpha_j = 1$ for normalization. It is worth noting that given the diversity of different pollutants when they are combined for impact assessment, a min-max scaling is employed for each type of pollutant, implying that $p_{ij} \in \left[0, 1\right]$ and $w_i \in \left[0, l_i\right]$ which can be seen as a parameter proportionally weighted based on the length of the $i$'th edge road segment $l_i$ and the vulnerability of the cyclist to the $K$ types of pollutants. This further implies that if the cyclist's vulnerability across different road segments is the same, the optimization model should favour the segment with the shortest length. 

Given this context, our optimization problem is to find a path $p = \langle v_1, v_2, \dots, v_h\rangle$ which consists of a sequence of vertices starting from the source vertex $v_1 \in V$ and terminating at the destination vertex $v_h \in V$ that minimizes the sum of the weights of its constituent edges. Mathematically, we have:

\begin{equation} \label{spp}
    \underset{p}{\operatorname{argmin}} \sum_{i=1}^{h-1} w(v_i, v_{i+1})
\end{equation}

\noindent where $w(u, v) \in W$ denotes the weight of the edge connecting vertices $u$ and $v$. 
    
    \section{System Architecture}
    \label{sec: arch}
\begin{figure}[ht]
    \vspace{-0.1in}
    \centering
    \includegraphics[width=\linewidth]{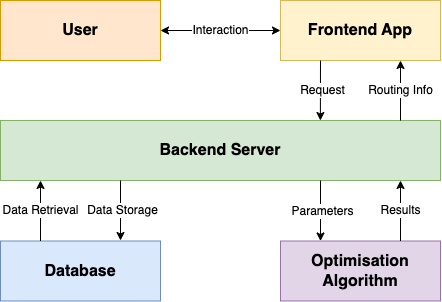}
    \caption{An overview of the system architecture}
    \vspace{-0.1in}
    \label{fig: overview}
\end{figure}

    This section introduces the proposed system architecture designed to address the research questions outlined in \autoref{sec: state}. An overview of the system architecture is presented in \autoref{fig: overview}. The diagram illustrates the components of our system, which consists of a mobile application serving as the frontend to interact with users, a backend server responsible for data transfer and processing, a database for data storage, and an optimisation algorithm to generate optimised routes based on input road parameters. Our project is open-source on GitHub\footnote{\url{https://github.com/SFIEssential/BreathGreen}}, so any interested readers are encouraged to visit and check out.

\subsection{Frontend}

    Our Android mobile application was developed using IntelliJ IDEA and subsequently tested on an Android Virtual Device featuring a $Google\,Play\,Intel\,Atom\,(x86)$ processor. The interactive map functionality within the application was implemented using the Google Maps API\footnote{\url{https://developers.google.com/maps/documentation/directions}}, which enables the display of origin and destination locations, as well as routes with the shortest distance or lowest pollutant index. 

\subsection{Backend}

    Based on a data-driven approach, for the purpose of real-time response, we utilised Flask \cite{grinberg2018flask}, a backend framework written in Python. Following the design of REST, the Flask server connects with frontend using JSON and calculates various metrics for self-optimisation. When starting the server, it loads the geographical information of Dublin from OpenStreetMap and appends the pollutant data collected from Google to generate a new digital map containing pollutant attributes on the top of roads. The routing algorithm that we propose calculates pollution level based on different setting acquired from users and output the entire route from start to the end with lowest degree of pollutants around. 

\subsection{Database \& Algorithm}

    The interaction between our database and the backend is bidirectional. Data received from the backend is stored in our database, and the database is also used to retrieve and return data to the backend as per its requirements.

    Similarly, the optimisation algorithm generates path recommendations based on the parameters received from the backend. The algorithm processes this information, calculate the optimised path and returns the results to the server. A more detailed explanation of the algorithm will be provided in the subsequent section.
    
    \section{Algorithm \& Experiments}
    \label{sec: algo_exp}
    \subsection{Algorithm}
    
    There are several well-known algorithms available to address \eqref{spp}, including Dijkstra's algorithm, Floyd-Warshall algorithm, and Bellman-Ford algorithm \cite{sapundzhi2018optimization}. These algorithms are designed to identify the shortest or most efficient path between two points in a weighted directed graph. For ease of reference, we shall now present a description of Dijkstra's Algorithm with reference to \cite{cormen2009introduction, Szczepanski2021} below. In \autoref{alg: dijkstra}, the vertex set is denoted by $Q$, and the array $dist$ contains the current distances from the starting point $s$ to other vertices, where $dist[v]$ represents the current distance from $s$ to vertex $u$. The array $prev$ contains pointers to the previous-hop nodes on the shortest path from $s$ to the given vertex, or equivalently, it is the next hop on the path from the given vertex to the source. The variable $alt$ denotes the length of the path from the root node to the neighbour node $v$ if it were to go through $u$. If this path is shorter than the current shortest path recorded for $v$, the current path would be replaced with this $alt$ path.
    
    \begin{algorithm}
        \caption{Dijkstra's Algorithm}
        \label{alg: dijkstra}
        \begin{algorithmic}
            \renewcommand{\algorithmicrequire}{\textbf{Input:}}
            \renewcommand{\algorithmicensure}{\textbf{Output:}}
            \REQUIRE $G=(V,E,W)$, a weighted directed graph, and $s, d \in V$, the starting point $s$ and destination point $d$
            \ENSURE  $p$, the shortest path for \eqref{spp}
            \FORALL{vertex $v$ in graph $G$}
                \STATE $dist[v] \gets infinite$
                \STATE $prev[v] \gets undefined$
                \STATE Add vertex $v$ to vertex set $Q$
            \ENDFOR
            \STATE $dist[s] \gets 0$
            \WHILE{$Q$ is not empty}
                \STATE $u \gets \underset{i}{\operatorname{argmin}}\{dist[i]\,|\,\forall i \in Q\} $
                \STATE Remove $u$ from $Q$
                \FORALL{neighbour $v$ of $u \in Q$}
                    \STATE $alt \gets dist[u] + w(u,v)$
                    \IF{$alt<dist[v]$}
                        \STATE $dist[v] \gets alt$
                        \STATE $prev[v] \gets u$
                    \ENDIF
                \ENDFOR
            \ENDWHILE
            \STATE Set $p$ as the shortest path to $d$ using $prev$ array
        \end{algorithmic}
    \end{algorithm}

\subsection{Dataset}
    
    This section describes the \texttt{AirView{\_}Dublin{\_}City} (\texttt{ADC}) dataset which is recently released and the preprocessing steps followed to address challenges faced during the data collection stage. Additionally, we provide insights and visualizations of the resulting dataset.
    
    \textbf{Google Project Air View Data - Dublin City (May 2021 - August 2022):}
    This dataset was collected by Google and Dublin City Council as part of Project Air View Dublin. Google's first electric Street View car, equipped with Aclima’s mobile air sensing platform, drove through the roads of Dublin City measuring street-by-street air quality. The data collection process predominantly took place from Monday to Friday between 9:00 am and 5:00 pm from May 2021 to August 2022. Consequently, the dataset represents typical daytime weekday air quality. The sensors on the car measured pollution on each street and highway at 1-second intervals while driving with the flow of traffic at normal speeds. The pollutants included in the dataset are Carbon Monoxide (${\rm CO}$), Carbon Dioxide (${\rm CO_2}$), Nitrogen Dioxide (${\rm NO_2}$), Nitric Oxide (${\rm NO}$), Ozone (${\rm O_3}$), and Particulate Matter ${\rm PM_{2.5}}$ (including size-resolved particle counts from 0.3 - 2.5 \textmu m). The following two versions of the dataset are publicly available\footnote{\url{https://data.smartdublin.ie/dataset/google-airview-data-dublin-city}}:
    
    \begin{itemize}
        \item \texttt{Airview{\_}Dublin{\_}City{\_}Measurements} (\texttt{ADC-M}) is the 1-second intervals data captured during the period.
        \item \texttt{AirView{\_}Dublin{\_}City{\_}RoadData} (\texttt{ADC-R}) is the 1-second data points aggregated in approximately 50m road segments.
    \end{itemize}
    
    In this work, the \texttt{ADC-R} was used because it groups the individual points/measurements from \texttt{ADC-M} into segments. To report the concentration of each pollutant per segment, first, they get the mean of the number of measurements on this road segment, then count the number of drive passes on this road segment. Finally, for each pollutant, they report the concentration (median of drive pass mean) in µg/m3. \texttt{ADC-R} also contains the OSM{\_}IDs segments to associate this dataset with the \texttt{OpenStreetMap} (\texttt{OSM}) dataset. The OSM{\_}IDs nodes were extracted and fed into the optimization algorithm. 

\subsection{Data Cleaning \& Prepossessing}

\begin{figure*}[ht]
        \vspace{-0.1in}
        \centering
        \subfigure[3 neighbours.]{
            \includegraphics[width=0.31\textwidth]{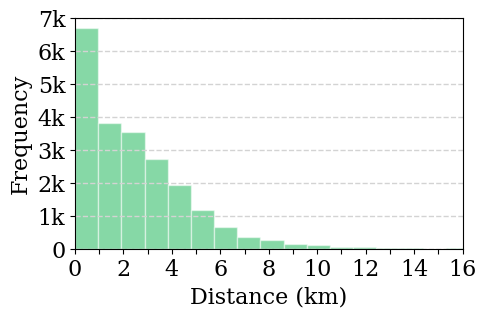}
            \label{subfig: n3}
        }
        \subfigure[5 neighbours.]{
            \includegraphics[width=0.31\textwidth]{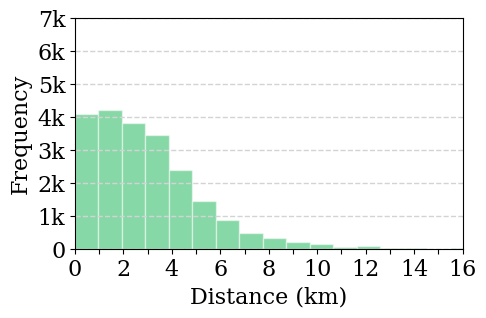}
            \label{subfig: n5}
        }
        \subfigure[10 neighbours.]{
            \includegraphics[width=0.31\textwidth]{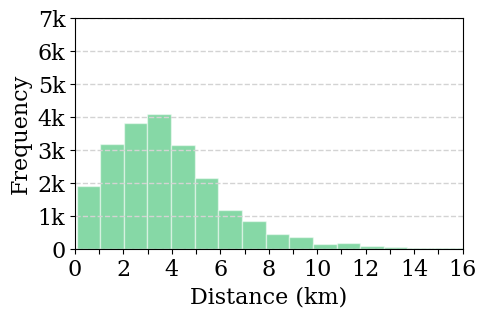}
            \label{subfig: n10}
        }
        \caption{Histogram of the distances to the neighbour segments used to fill the missing values for three cases: 3, 5, and 10 neighbours.}
        \vspace{-0.1in}
        \label{fig: neighbour hist}
    \end{figure*}

\begin{figure}[ht]
        \centering
        \includegraphics[width=\linewidth]{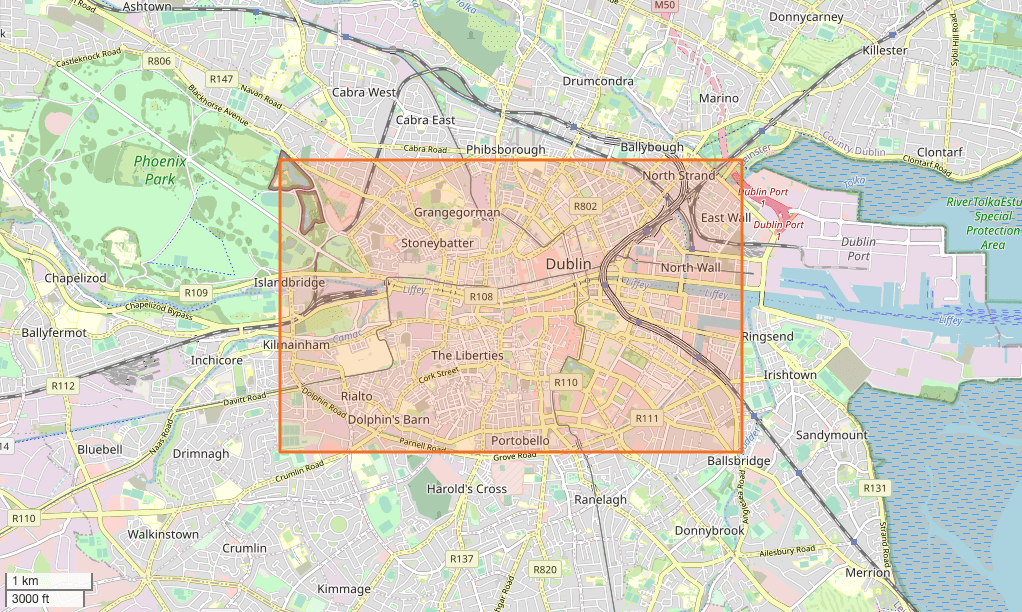}
        \caption{Dublin area of interest}
        \label{fig: snapshot_dublin}
    \end{figure}

    In this work, the \texttt{ADC-R} dataset was used as mentioned previously. Even though this dataset is an aggregation of the raw data \texttt{ADC-M} and was already processed by Google, this work required the following pre-processing:
    
    \begin{itemize}
        \item Data interpolation for missing and negative values: the dataset includes NaNs and negative values, which may arise from sensor malfunctions or miscalibrations, potentially impacting the optimization algorithm. To address this issue, a linear interpolation method was applied to each pollutant column. Initially, the dataset was sorted by OSMID, and all negative values were replaced with NaNs. Subsequently, a linear interpolation was performed using the Python Pandas library, enabling the estimation of missing values and the smoothing of the dataset.

        \item Cross-referencing of the \texttt{OSM} and \texttt{ADC-R} datasets: the optimisation algorithm utilizes the nodes and segments from the \texttt{OSM} dataset to determine the greener path. To establish a link between these two datasets, the OSM\_IDs are used as a reference. It is important to note that the vehicle used to collect the Google Air Quality dataset did not traverse all streets and roads in Dublin, resulting in measurements being available only for specific segments. To assign these measurements to the corresponding segments in the \texttt{OSM} dataset, we conducted the matching process by utilising the OSM\_ID column to perform a matching process between the road dataset and the \texttt{OSM}. Segments existing in both datasets were selected and their corresponding pollutant measurements were merged. This resulted in the creation of a final dataset with the following format: [osm\_id, NO2\_ugm3, NO\_ugm3, CO2\_mgm3, CO\_mgm3, O3\_ugm3, PM25\_ugm3]. The optimization algorithm is applied to this final dataset.
            
        \item Furthermore, in order to address the issue of missing data on particular segments, a clustering algorithm is employed to identify the nearest neighbours from those segments. In particular, the position of a road segment is represented by its edge points, encompassing two endpoints and, if present, all inflexion points. Subsequently, the BallTree algorithm is applied to select the $n$ closest neighbours for each target, which refers to the edge point in a road segment with missing data, within our dataset. The attributes of the target are then determined by calculating the mean value of its neighbours' attributes, thereby enabling the missing data to be filled in based on geographic relevance. The histogram depicted in \autoref{fig: neighbour hist} provides a visualisation of the distances between edge points with missing data and their closest neighbours. Considering the observed distribution of distances, the selection of 3, 5, or 10 neighbours is made. With a distance threshold of 3 km being assigned to limit the longest distance between a specific target to its neighbours, the fraction of the interpolated segments with their value assigned using the 3, 5 and 10 closest neighbours are 34.10\%,  30.84\% and 23.95\% respectively. In this study, we selected 5 as the neighbour number because it provides a good filling rate with an acceptable number of supportive neighbours for each road segment.

        \item Finally, we delimited the area of interest with a focus on the central region of Dublin. In order to cover the areas frequented by shared bikes, we define the area of interest based on the map of NOW dublinbikes\footnote{\url{https://www.dublinbikes.ie/en/mapping}} and all segments and nodes within the coordinates [-6.230852, 53.359967, -6.310015, 53.330091] (as shown in \autoref{fig: snapshot_dublin}) were extracted.
    \end{itemize}

\begin{figure}[ht]
        \centering
        \includegraphics[width=\linewidth]{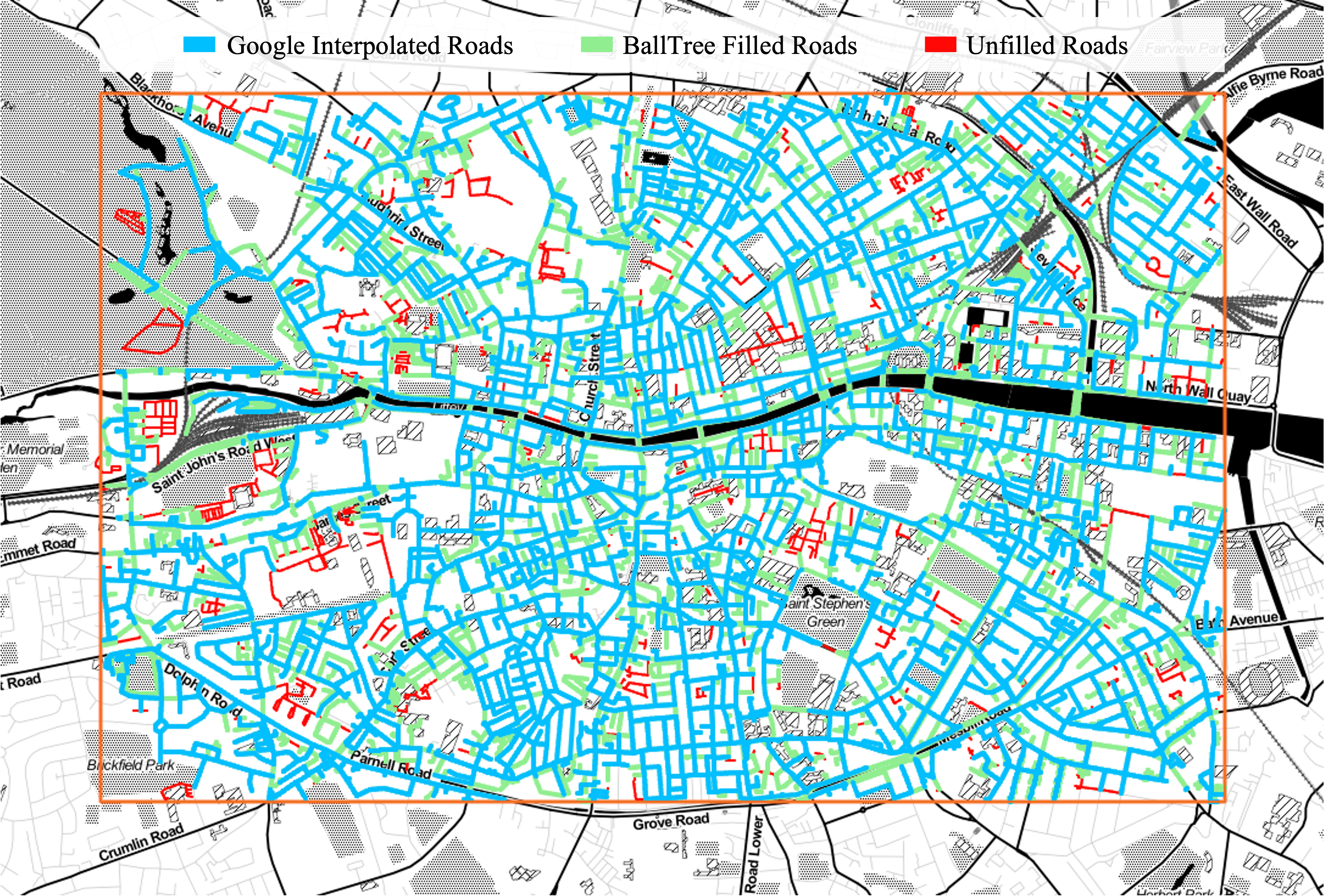}
        \caption{Data before and after cleaning and preprocessing.}
        \vspace{-0.1in}
        \label{fig: data filling result}
    \end{figure}

    The data before and after cleaning and preprocessing is visually depicted on the map in \autoref{fig: data filling result}, where the orange boundary corresponds to the area shown in \autoref{fig: snapshot_dublin} and contains 13,697 road segments in total. The blue routes on the map, with a number of 7,710, represent the road segments with Google interpolations in the original dataset, the green routes, with a number of 4,224, represent the roads interpolated by our filling process and the red routes, with a number of 1,763, are the roads still with missing values even after the filling process. In one word, the data missing rate has decreased from 43.71\% to 12.87\% due to our data-filling process.
    
\subsection{Green Route Planning}

    Our experiments were designed to assess the impact of our algorithm on the selection of green routes in various real-world scenarios. To achieve this, we generated and compared the shortest routes with the greenest routes under different setups, as outlined below:
    
    \begin{enumerate}
    
        \item In the first setup, we considered all pollutants and assigned them equal weights. We conducted the experiments with a thousand randomly selected origin-destination pairs on the map, focusing primarily on walking and private micromobility modes such as private bicycles and e-bikes. Thus, we denote this setup as ``\textbf{Private Device}''.
        
        \item In the second setup, we again considered all pollutants and assigned them equal weights. However, this time we conducted the experiments with ten origin-destination pairs, selecting GPS coordinates from the map of NOW dublinbikes. This setup specifically targeted transportation systems with shared devices, such as shared bicycles and shared e-bikes, and we call this setup ``\textbf{Shared Device}''.

        \item In the third setup, we examined the individual influence of different pollutants. We conducted the experiments with one origin-destination pair selected on the map, varying the weights assigned to different pollutants. For example, we assigned a weight of 1 to one pollutant, such as ${\rm CO}$, and a weight of 0 to the other pollutants. This allowed us to obtain green routes that focused solely on reducing the specific pollutant. This setup aimed to evaluate the performance of our optimization algorithm under different pollutant weightings. This setup is named ``\textbf{Single Pollutant}''.
        
    \end{enumerate}

    By conducting these experiments, we aimed to gain insights into the performance and effectiveness of our optimization algorithm in different real-world scenarios and pollutant weightings. The results and relevant discussions are included in \autoref{sec: res_dis}.


    \section{Results \& Discussion}
    \label{sec: res_dis}
    In this section, we present a summary of our experimental results and provide visual representations. Furthermore, we offer detailed discussions and analysis of the obtained results.

\subsection{Private Device}

\begin{figure}[ht]
        \centering
        \includegraphics[width=\linewidth]{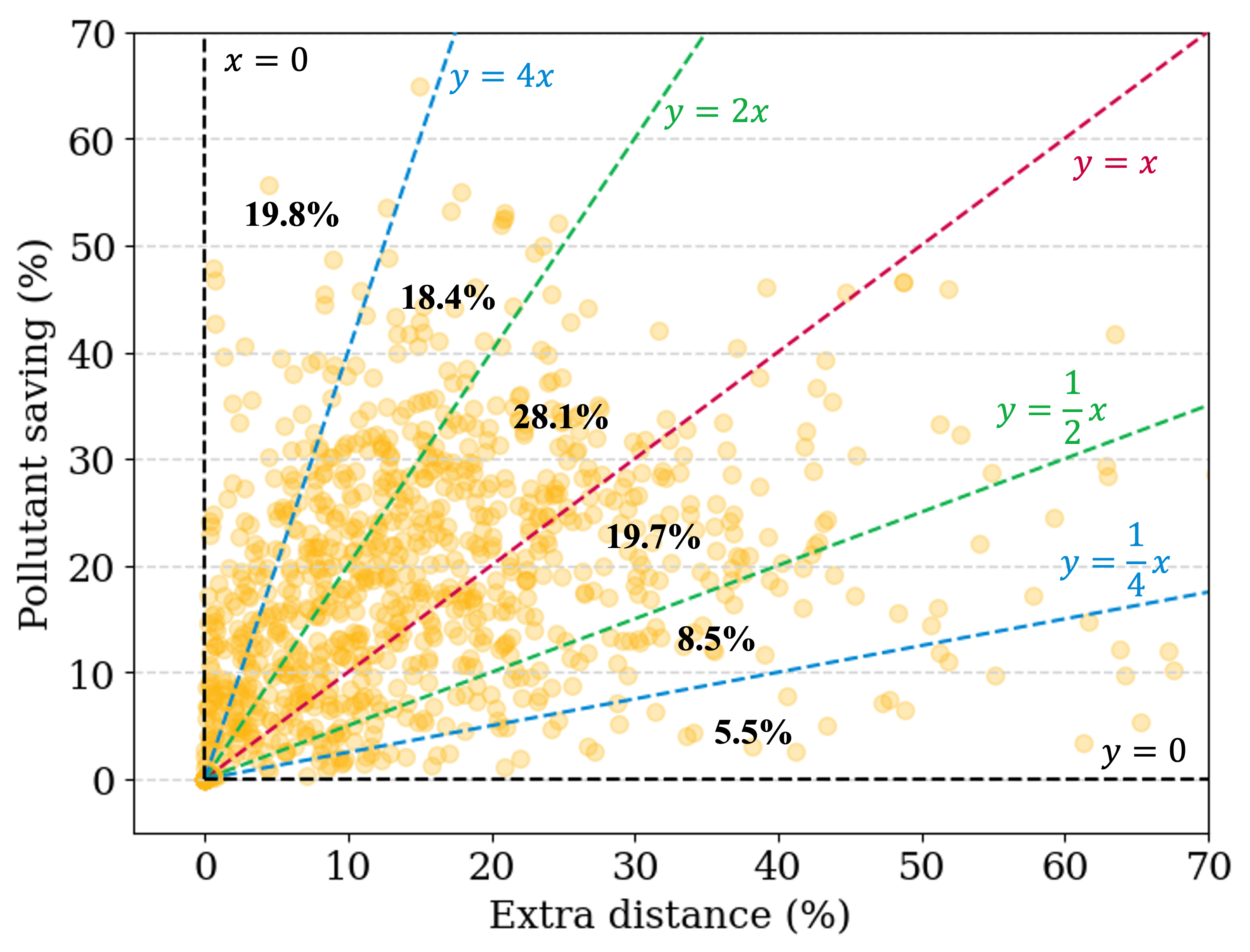}
        \caption{Pollutant saving and distance increase.}
        \label{fig: p over d}
    \end{figure}

    We performed a series of experiments with a thousand different origin-destination pairs, and the summarised results are presented in \autoref{tab: private device results}. The findings clearly indicate that while the green path shows an average increase in trip distance of 15.15\% compared to the shortest route, it simultaneously leads to a reduction in pollutant intake by 17.87\%. \autoref{fig: p over d} visualised the relationship between the pollutant saving rate and distance increase rate, with yellow scatter points representing the data. The graph is partitioned into six regions based on the linear boundaries defined by the equations $y=0$, $y=0.25x$, $y=0.5x$, $y=x$, $y=2x$, $y=4x$ and $x=0$. Additionally, the percentage of scatter points within each region is presented on the graph. The analysis of the graph reveals that the majority of scatter points are located in the region bounded by $y=x$ and $y=2x$. This observation indicates that, in most cases, the rate of pollutant saving exceeds the rate of distance increase. Hence, our algorithm demonstrates a favourable trade-off between reducing pollutant inhalation and the increase in trip distance. 
    
    It is worth noting that in some cases, certain routes in our experiments coincide with both the shortest path and the green path, resulting in zero values for pollutant savings and additional distance.

    \begin{table}
        \caption{Private device results.}
        \label{tab: private device results}
        \begin{tabularx}{\linewidth}{@{\extracolsep{\fill}}c c c c c}
            \toprule
            & \makecell[c]{ShortPath \\ distance (m)} & \makecell[c]{GreenPath \\ distance (m)} & \makecell[c]{Pollutant \\ saving (\%)} & \makecell[c]{Extra \\ distance (\%)} \\
            \midrule
            mean    & 2870.31 & 3328.76 & 17.87 & 15.15 \\
            std     & 1267.07 & 1572.03 & 12.20 & 13.93  \\
            min     & 208.20  & 208.20  & 0     & 0     \\
            25\%    & 1927.26 & 2137.51	& 7.72  & 5.09  \\
            50\%    & 2820.44 & 3284.45 & 17.20 & 12.11 \\
            75\%    & 3711.74 & 4304.19 & 26.43 & 21.43 \\
            max     & 7303.63 & 9265.65 & 64.99 & 83.22 \\
            \bottomrule
        \end{tabularx}
    \end{table}

\subsection{Shared Device}

    Another series of experiments were conducted with ten different origin-destination pairs, and the results are summarised in \autoref{tab: shared device results}. Similar findings can be observed in this set of experiments compared to the previous ones. Our algorithm, when applied to shared micro-mobilities, also leads to reduced pollutant inhalation at the expense of a certain increase in trip distance.

    \begin{table}
        \caption{Shared device results.}
        \label{tab: shared device results}
        \begin{tabularx}{\linewidth}{@{\extracolsep{\fill}}c c c c c}
            \toprule
            & \makecell[c]{ShortPath \\ distance (m)} & \makecell[c]{GreenPath \\ distance (m)} & \makecell[c]{Pollutant \\ saving (\%)} & \makecell[c]{Extra \\ distance (\%)} \\
            \midrule
            mean    & 2460.99 & 2921.88 & 14.68 & 17.59 \\
            std     & 1206.03 & 1459.44 & 10.37 & 17.21 \\
            min     & 769.17  & 769.17  & 0     & 0     \\
            25\%    & 1480.41 & 1652.79	& 5.93  & 4.94  \\
            50\%    & 2413.75 & 2762.74 & 15.42 & 15.62 \\
            75\%    & 3464.61 & 4253.95 & 20.84 & 21.81 \\
            max     & 4264.24 & 5079.85 & 31.12 & 54.36\\
            \bottomrule
        \end{tabularx}
    \end{table}

\subsection{Single Pollutant}

\begin{figure*}[ht]
        \vspace{-0.1in}
        \centering
        \subfigure[Paths determined by single pollutant.]{
            \includegraphics[width=\textwidth]{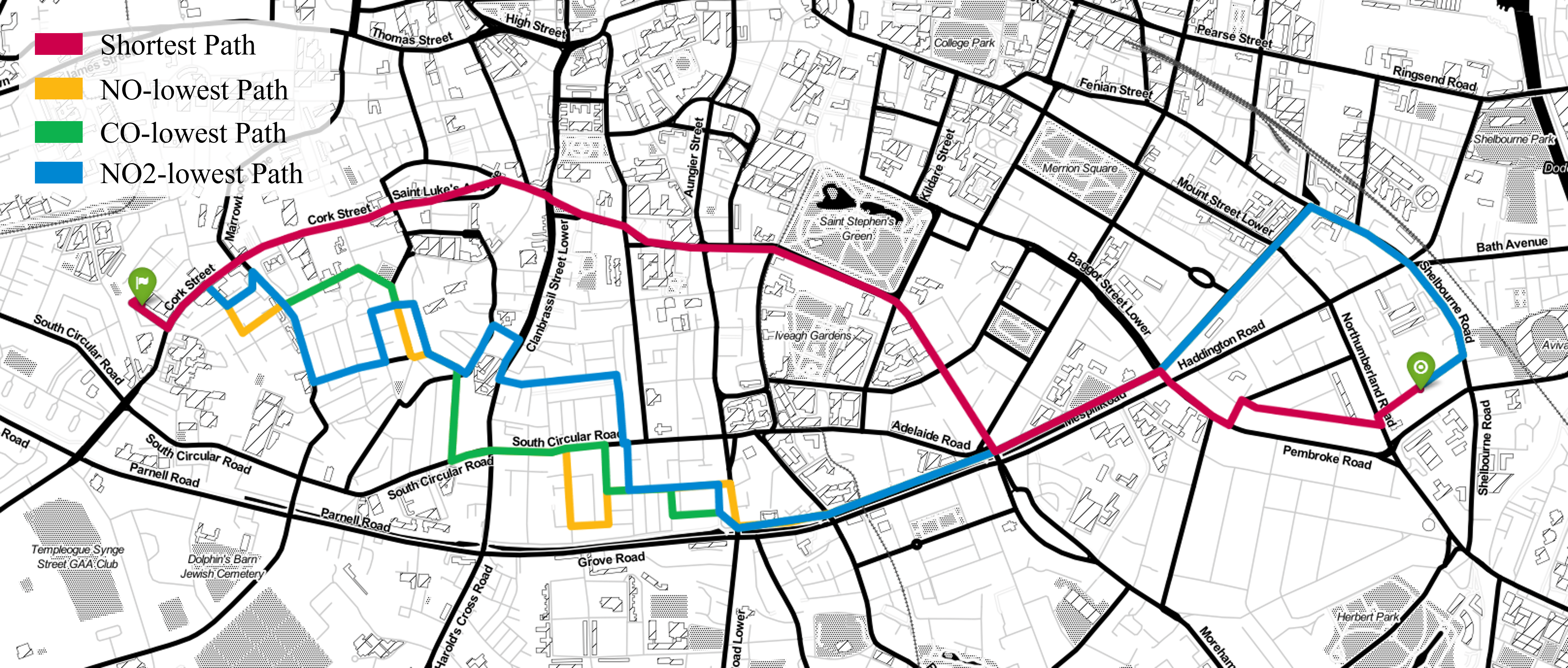}
            \label{subfig: single pollutant}
            \vspace{-0.1in}
        }
        \subfigure[${\rm NO_2}$ distribution.]{
            \includegraphics[width=\textwidth]{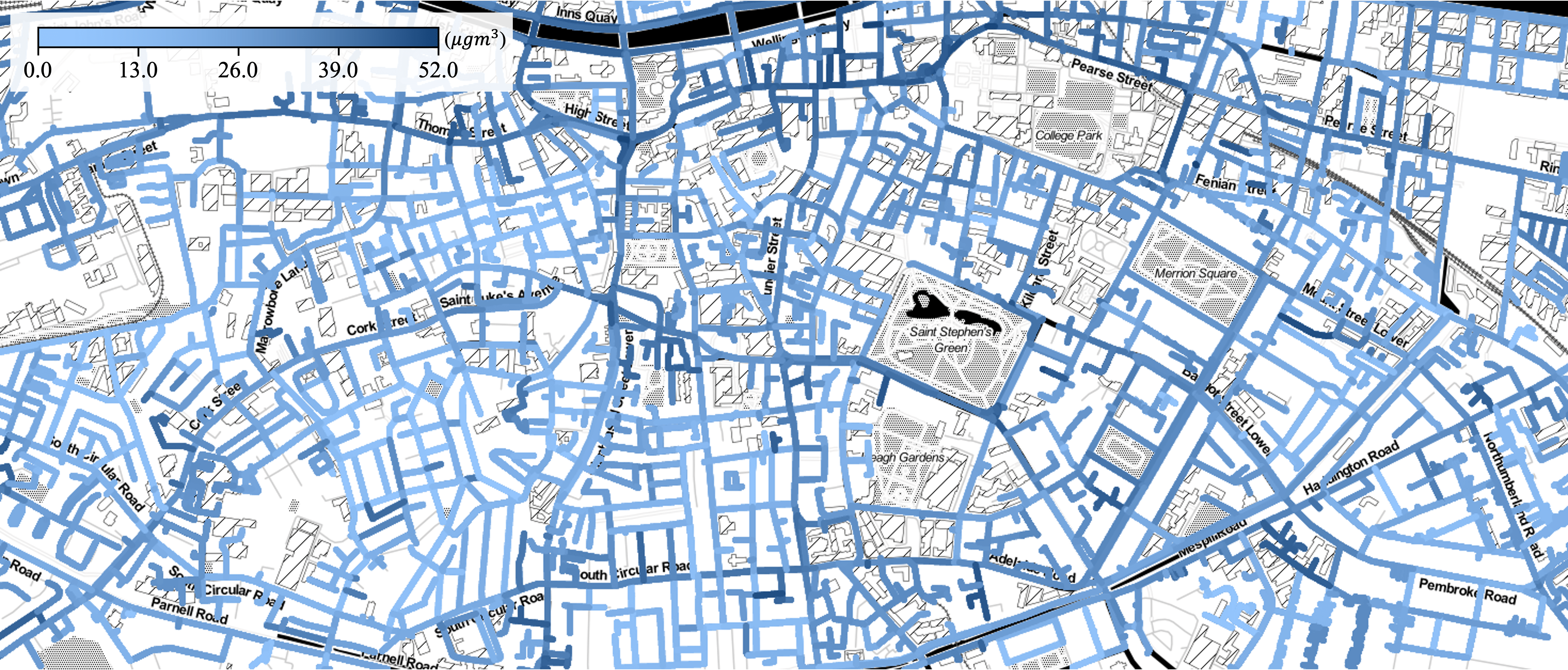}
            \label{subfig: no2}
        }
        \caption{Single pollutant results.}
        \vspace{-0.1in}
        \label{fig: single results}
    \end{figure*}

    The paths for different pollutants are depicted in \autoref{subfig: single pollutant}, with the red path indicating the shortest route, and the yellow, green, and blue paths determined by ${\rm NO}$, ${\rm CO}$, and ${\rm NO_2}$, respectively, resulting in pollutant reduction percentages of 52.07\%, 27.52\%, and 50.57\%. The values of each pollutant for different paths, as presented in \autoref{tab: single results}, demonstrate the algorithm's capability to recommend diverse routes for the same origin-destination pair based on varying pollutant weights, effectively reducing pollutant levels. For instance, taking ${\rm NO_2}$ as an example, the segment values around the trip trajectory, visualized in \autoref{subfig: no2}, indicate that the blue path in \autoref{subfig: single pollutant} exhibits the lowest ${\rm NO_2}$ levels.

    \begin{table*}[ht]
        \caption{Single pollutant descriptive statistics.}
        \label{tab: single results}
        \begin{tabularx}{\linewidth}{@{\extracolsep{\fill}}c c c c c c c}
            \toprule
            & \textbf{length\_m} & \textbf{CO\_mgm3} & \textbf{NO\_ugm3} & \textbf{NO2\_ugm3} & \textbf{O3\_ugm3} & \textbf{PM25\_ugm3}  \\
            \midrule
            Shortest path    & 4556.73 & 34.16 & 1648.08 & 1585.60 & 3900.73 & 647.14 \\
            ${\rm CO}$-lowest path   & 5643.65 & 26.88 & 853.61  & 904.94  & 3680.48 & 485.35 \\
            ${\rm NO}$-lowest path   & 5781.24 & 31.16 & 790.19  & 951.88  & 4358.15 & 557.70 \\
            ${\rm NO_2}$-lowest path  & 6127.13 & 32.30 & 1055.97 & 786.45  & 4573.79 & 588.11 \\
            ${\rm O_3}$-lowest path   & 5572.22 & 27.14 & 1053.94 & 1039.73 & 3456.92 & 475.91 \\
            ${\rm PM_{2.5}}$-lowest path & 5572.31 & 27.16 & 1009.75 & 989.46  & 3566.17 & 472.32 \\
            \bottomrule
        \end{tabularx}
    \end{table*}

    \section{Conclusions}
    \label{sec: con}
    Our research aimed to maximise health and environmental benefits for active transportation participants. We developed a system that recommends trip routes to minimise pollutant inhalation based on the analysis of air quality data in Dublin, Ireland, which was recently released by Google. Through various experiments, we evaluated the performance of our optimisation algorithm and found it to be reliable and practical in diverse real-world scenarios, catering to both private and shared micro-mobility users, such as e-bike riders and shared bike users. Our results lead to a remarkable reduction in pollutant intake, with a significant decrease of 17.87\% on average and a maximum of around 64.99\%. These findings underscore the efficacy and value of our approach in improving the environmental well-being of individuals. 

Furthermore, the enhancement of our work can be achieved through the implementation of a more sophisticated and rigorous pollutant evaluation standard or metrics, as well as by augmenting the completeness and comprehensiveness of the dataset. Additional avenues for improvement include conducting comparisons of pollution savings against estimated variations, such as day-to-day differences, peak and off-peak measurements, or disparities between routes in close proximity. Another viable approach entails modifying the 'least-polluting' algorithm by introducing penalties for specific turns at intersections or incorporating diverse travel speeds based on the road segments traversed and the presence of street lights. Addressing these areas of refinement holds the potential to bolster the quality and reliability of our research, thereby contributing to a more advanced and comprehensive understanding of the environmental impact of outdoor activities and the effectiveness of our system.

    \section*{Acknowledgement}
       \label{sec: ack}
            This research was conducted with the financial support of Science Foundation Ireland \textit{21/FFP-P/10266} and \textit{12/RC/2289\_P2} at Insight the SFI Research Centre for Data Analytics at Dublin City University. For the purpose of Open Access, the author has applied a CC BY public copyright licence to any Author Accepted Manuscript version arising from this submission.
    
    \bibliographystyle{ieeetr}
    \bibliography{references}

\begin{thebibliography}{10}

\bibitem{Elford2019}
S.~Elford and M.~D. Adams, ``Exposure to ultrafine particulate air pollution in
  the school commute: Examining low-dose route optimization with
  terrain-enforced dosage modelling,'' {\em Environmental Research}, vol.~178,
  p.~108674, Nov. 2019.

\bibitem{An2022}
F.~An, J.~Liu, W.~Lu, and D.~Jareemit, ``Comparison of exposure to
  traffic-related pollutants on different commuting routes to a primary school
  in jinan, china,'' {\em Environmental Science and Pollution Research},
  vol.~29, pp.~43319--43340, Jan. 2022.

\bibitem{Islam2021}
M.~A. Islam and Y.~Gajpal, ``Optimization of conventional and green vehicles
  composition under carbon emission cap,'' {\em Sustainability}, vol.~13,
  p.~6940, June 2021.

\bibitem{Abdullahi2021}
H.~Abdullahi, L.~Reyes-Rubiano, D.~Ouelhadj, J.~Faulin, and A.~A. Juan,
  ``Modelling and multi-criteria analysis of the sustainability dimensions for
  the green vehicle routing problem,'' {\em European Journal of Operational
  Research}, vol.~292, pp.~143--154, July 2021.

\bibitem{Olgun2021}
B.~Olgun, {\c{C}}.~Ko{\c{c}}, and F.~Alt{\i}parmak, ``A hyper heuristic for the
  green vehicle routing problem with simultaneous pickup and delivery,'' {\em
  Computers {\&} Industrial Engineering}, vol.~153, p.~107010, Mar. 2021.

\bibitem{Cai2021}
L.~Cai, W.~Lv, L.~Xiao, and Z.~Xu, ``Total carbon emissions minimization in
  connected and automated vehicle routing problem with speed variables,'' {\em
  Expert Systems with Applications}, vol.~165, p.~113910, Mar. 2021.

\bibitem{Sweeney2019}
S.~Sweeney, R.~Ordonez-Hurtado, F.~Pilla, G.~Russo, D.~Timoney, and R.~Shorten,
  ``A context-aware e-bike system to reduce pollution inhalation while
  cycling,'' {\em {IEEE} Transactions on Intelligent Transportation Systems},
  vol.~20, pp.~704--715, Feb. 2019.

\bibitem{Herrmann2018}
A.~Herrmann, M.~Liu, F.~Pilla, and R.~Shorten, ``A new take on protecting
  cyclists in smart cities,'' {\em {IEEE} Transactions on Intelligent
  Transportation Systems}, vol.~19, pp.~3992--3999, Dec. 2018.

\bibitem{Dong2019}
E.~Dong, L.~Zhang, and H.~Shi, ``Multi-objective optimization of hybrid
  electric vehicle powertrain,'' in {\em 2019 {IEEE} 3rd Information
  Technology, Networking, Electronic and Automation Control Conference
  ({ITNEC})}, {IEEE}, Mar. 2019.

\bibitem{Hong2021}
W.~Hong, I.~Chakraborty, H.~Wang, and G.~Tao, ``Co-optimization scheme for the
  powertrain and exhaust emission control system of hybrid electric vehicles
  using future speed prediction,'' {\em {IEEE} Transactions on Intelligent
  Vehicles}, vol.~6, pp.~533--545, Sept. 2021.

\bibitem{Umezawa2022}
Y.~Umezawa, K.~Yamauchi, H.~Seto, T.~Imamura, and T.~Namerikawa, ``Optimization
  of fuel consumption and {NO\textsubscript{x}} emission for mild {HEV} via
  hierarchical model predictive control,'' {\em Control Theory and Technology},
  vol.~20, pp.~221--234, May 2022.

\bibitem{Langbridge2022}
A.~Langbridge, P.~Ferraro, and R.~Shorten, ``Respiratory aware routing for
  active commuters,'' 2022.

\bibitem{LuengoOroz2019}
J.~Luengo-Oroz and S.~Reis, ``Assessment of cyclists' exposure to ultrafine
  particles along alternative commuting routes in edinburgh,'' {\em Atmospheric
  Pollution Research}, vol.~10, pp.~1148--1158, July 2019.

\bibitem{Jereb2018}
B.~Jereb, T.~Batkovi{\v{c}}, L.~Herman, G.~{\v{S}}ipek, {\v{S}}.~Kov{\v{s}}e,
  A.~Gregori{\v{c}}, and G.~Mo{\v{c}}nik, ``Exposure to black carbon during
  bicycle commuting{\textendash}alternative route selection,'' {\em
  Atmosphere}, vol.~9, p.~21, Jan. 2018.

\bibitem{Sweeney2017}
S.~Sweeney, R.~Ordonez-Hurtado, F.~Pilla, G.~Russo, D.~Timoney, and R.~Shorten,
  ``Cyberphysics, pollution mitigation, and pedelecs,'' {\em ArXiv}, 06 2017.

\bibitem{Samal2020}
K.~K.~R. Samal, K.~S. Babu, and S.~K. Das, ``{ORS}: The optimal routing
  solution for smart city users,'' in {\em Lecture Notes in Electrical
  Engineering}, pp.~177--186, Springer Singapore, 2020.

\bibitem{Samal2021}
K.~Samal, K.~Babu, and S.~Das, ``Predicting the least air polluted path using
  the neural network approach,'' {\em {ICST} Transactions on Scalable
  Information Systems}, vol.~8, p.~170250, Oct. 2021.

\bibitem{Gu2018}
Y.~Gu, M.~Liu, M.~Souza, and R.~N. Shorten, ``On the design of an intelligent
  speed advisory system for cyclists,'' in {\em 2018 21st International
  Conference on Intelligent Transportation Systems ({ITSC})}, {IEEE}, Nov.
  2018.

\bibitem{Wu2022}
T.-G. Wu, Y.-D. Chen, B.-H. Chen, K.~H. Harada, K.~Lee, F.~Deng, M.~J. Rood,
  C.-C. Chen, C.-T. Tran, K.-L. Chien, T.-H. Wen, and C.-F. Wu, ``Identifying
  low-{PM}2.5 exposure commuting routes for cyclists through modeling with the
  random forest algorithm based on low-cost sensor measurements in three asian
  cities,'' {\em Environmental Pollution}, vol.~294, p.~118597, Feb. 2022.

\bibitem{grinberg2018flask}
M.~Grinberg, {\em Flask web development: developing web applications with
  python}.
\newblock " O'Reilly Media, Inc.", 2018.

\bibitem{sapundzhi2018optimization}
F.~Sapundzhi and M.~Popstoilov, ``Optimization algorithms for finding the
  shortest paths,'' {\em Bulgarian Chemical Communications}, vol.~50,
  no.~Special Issue B, pp.~115--120, 2018.

\bibitem{cormen2009introduction}
T.~H. Cormen, C.~E. Leiserson, R.~L. Rivest, and C.~Stein, {\em Introduction to
  Algorithms}.
\newblock MIT Press, 3~ed., 2009.

\bibitem{Szczepanski2021}
R.~Szczepanski and T.~Tarczewski, ``Global path planning for mobile robot based
  on artificial bee colony and dijkstra's algorithms,'' in {\em 2021 {IEEE}
  19th International Power Electronics and Motion Control Conference ({PEMC})},
  {IEEE}, Apr. 2021.

\end{thebibliography}
\end{document}